\documentclass{aastex62}

\shorttitle{A Comparative Study of Data-based MHD Simulations}
\shortauthors{Inoue et al.}
\begin{document}

\title{A Comparative Study of  Solar Active Region 12371 with Data-constrained  and Data-driven MHD Simulations}

\correspondingauthor{Satoshi Inoue}
\email{Satoshi.Inoue@njit.edu}

\author[0000-0001-5121-5122]{Satoshi Inoue}
\affiliation{Center for Solar-Terrestrial Research, New Jersey Institute of Technology,  
             University Heights, Newark, NJ 07102-1982, USA}

\author[0000-0001-9046-6688]{Keiji Hayashi}
\affiliation{George Mason University, 4400 University Dr, Fairfax, VA 22030, USA}

\author[0000-0002-4675-4460]{Takahiro Miyoshi}
\affiliation{Graduate School of Advanced Science and Engineering, Hiroshima University, 1-3-1 Kagamiyama, Higashihiroshima 739-8526, Japan}

\author[0000-0002-8179-3625]{Ju Jing}
\affiliation{Center for Solar-Terrestrial Research, New Jersey Institute of Technology,  
University Heights, Newark, NJ 07102-1982, USA}

\author[0000-0002-5233-565X]{Haimin Wang}
\affiliation{Center for Solar-Terrestrial Research, New Jersey Institute of Technology,  
             University Heights, Newark, NJ 07102-1982, USA}



 \begin{abstract}
 % ==================================================================================================================================================
We performed two data-based magnetohydrodynamic (MHD) simulations for solar active region 12371 which  produced an M6.5 flare. The first simulation is a full data-driven simulation where the initial condition is given by a nonlinear force-free field (NLFFF). This NLFFF was extrapolated from  photospheric magnetograms approximately 1 hour prior to the flare, and then a time-varying photospheric magnetic field is imposed at the bottom surface. The second simulation is also a data-driven simulation, but it stops driving at the bottom before the time of flare onset and then switches to the data-constrained simulation, where the horizontal component of the magnetic field varies according to an induction equation while the normal component is fixed with time. Both simulations lead to an eruption, with both simulations producing highly twisted field lines before the eruption which were not found in the NLFFF alone. After the eruption, the first simulation based on the time-varying photospheric magneitic field, continues to produce sheared field lines after the flare without reproducing phenomena such as post-flare loops. The second simulation reproduces the phenomena associated with flares well. However in this case the evolution of the bottom magnetic field is inconsistent with the evolution of the observed magnetic field. In this letter, we report potential advantages and disadvantages in data-constrained and data-driven MHD simulations that need to be taken into consideration by future studies.
 
\end{abstract}

%\keywords{Sun:magnetic field --- Sun:solar flares --- Sun:coronal mass ejections --- surveys}

% ================================================================================================================================================
    \section{Introduction} \label{sec:intro}
% ================================================================================================================================================
The evolution and eruption of the solar magnetic fields is one of long-standing problems in solar physics, and has been thoroughly investigated using both observational and numerical approaches ({\it e.g.,} \citealt{Shibata2011}, \citealt{Toriumi2019}). Although traditionally observational and simulation studies have been conducted near independently, recent state-of-the-art solar satellites and ground-based observations enable us to combine them. For example, a nonlinear force-free field (NLFFF) extrapolation calculated from an observed photospheric magnetic field makes it possible to show the three-dimensional (3D) magnetic field before the onset of solar flares (\citealt{Inoue2016}). Since the NLFFF is extrapolated under a force-free approximation, highly twisted field lines, which accumulate the free magnetic energy to produce the flare, are reproduced well along the polarity inversion line (PIL) ({\it e.g.}, \citealt{Schrijver2008}). This result cannot be obtained using a potential magnetic field model. Tracing the temporal evolution of the NLFFF is a powerful tool for us to understand the magnetic field leading up to the eruption, and to allow comparison of the fields before and after the flare (\citealt{Sun2012}, \citealt{Jiang2014}). However, these modes provide only a single snapshot in the evolution process, and the extrapolated magnetic fields are bounded to the force-free condition. 
 
 To clarify the dynamics during flaring, data-based magnetohydrodynamics (MHD) simulations have been conducted in recent years. There are two main approaches, one is the data-constrained MHD simulation ({\it e.g.,} \citealt{Jiang2013}, \citealt{Inoue2014}, \citealt{Muhamad2017}), where the initial condition is taken from  observations, but the subsequent calculation of the magnetic field no longer follows these observations. The second is a more advanced approach called “data-driven” simulation (see more recent review \citealt{Jiang2022}) that keeps the magnetic field in every step constrained by the observations. In many data-constrained MHD simulations, the NLFFF is given as the initial condition. Although the dynamics are allowed to be free from the force-free restriction, the tangential component of the bottom magnetic field is inconsistent with the observations while the normal one is fixed in time in many cases. Nevertheless, the data-constrained simulations have reproduced observed phenomena associated with flares ({\it e.g.}, \citealt{Inoue2014}, \citealt{Inoue2015}). Data-driven simulations drive the coronal magnetic field based on a time-series of observed photospheric magnetic fields or a time-series of velocity or electric fields, both of which are derived from the time-series of photospheric magnetic field observations. Several simulations support the observational findings (\citealt{Cheung2012}, \citealt{Jiang2016}, \citealt{Hayashi2018}, \citealt{Kaneko2021}, \citealt{Kilpua2021}). \cite{Leake2017} and \cite{Inoue2022} and show good performance within their data-driven simulation framework. However, using ground-truth data, \cite{Toriumi2020} shows that different data-driven models give different solutions. Thus, although many studies that include data-constrained and data-driven simulations have been conducted and have shown strengths for each method, there have been few discussions directly dealing with the potential advantages and disadvantages of each approach. 
 
 In this letter, we conduct an NLFFF extrapolation and data-based MHD simulations using both data-driven and data-constrained approaches to discuss the advantages and disadvantages of each method. The photospheric magnetic fields of solar active region (AR) 12371 were used for this study from 16:36 UT to 20:00 UT on June 15 2015 that covers build up to the M6.5 flare event which was reported in (\citealt{Wang2017}, \citealt{Kang2019}). Finally, we discuss what is required to further improve the reliability of data-based MHD simulations.

% ===============================================================================================================================================
    \section{Numerical Methods and Observations} 
% ===============================================================================================================================================
  We solve the following zero-beta MHD equations shown in \cite{Inoue2014} and \cite{Inoue2016},
  
% --------------------------------------------------------------------------
% Mass Equation
% --------------------------------------------------------------------------
  \begin{equation}
  \frac{\partial \rho}{\partial t} = -{\bf \nabla}\cdot(\rho {\bf v})+\zeta{\bf \nabla}^2(\delta \rho),
  \label{eq_mass}
  \end{equation}

% ------------------------------------------------------------------------
% Equation of Motion
% ------------------------------------------------------------------------
  \begin{equation}
  \frac{\partial {\bf v}}{\partial t} 
                        = - ({\bf v}\cdot {\bf \nabla}){\bf v}
                          + \frac{1}{\rho} {\bf J} \times {\bf B}
                          + \nu{\bf \nabla}^{2}{\bf v},
  \label{eq_motion}
  \end{equation}

% ------------------------------------------------------------------------
% Induction equation
% ------------------------------------------------------------------------
  \begin{equation}
  \frac{\partial {\bf B}}{\partial t} 
                        =  {\bf \nabla}\times({\bf v}\times{\bf B})
                        +  \eta {\bf \nabla}^2 {\bf B}
                        -  {\bf \nabla}\phi, 
  \label{induc_eq}
  \end{equation}

% ------------------------------------------------------------------------
% Ampere's low
% ------------------------------------------------------------------------
  \begin{equation}
  {\bf J} = {\bf \nabla}\times{\bf B},
  \end{equation}
  
% ------------------------------------------------------------------------
% Dedner 
% ------------------------------------------------------------------------
  \begin{equation}
  \frac{\partial \phi}{\partial t} + c^2_{h}{\bf \nabla}\cdot{\bf B} 
    = -\frac{c^2_{h}}{c^2_{p}}\phi,
  \label{div_eq}
  \end{equation}
  to conduct the NLFFF extrapolation and data-based MHD simulations where ${\bf B}$ is the magnetic flux density, ${\bf v}$ is the velocity, ${\bf J}$ is the electric current density, $\rho$ is the density, and $\phi$ is the convenient potential to remove errors derived from ${\bf \nabla}\cdot {\bf B}$ (\citealt{Dedner2002}). Furthermore, $\delta \rho$ is defined as $\rho-\rho_0$, where $\rho_0$ corresponds to the initial density. The length, magnetic field, density, velocity, and time are normalized by $L^{*}=3.45\times 10^{8}(m)$ , $B^{*}=0.28(T)$, $\rho^{*}=6.15\times 10^{-8}(kg/m^3)$ , $V_{\rm A}^{*}\equiv B^{*}/(\mu_{0}\rho^{*})^{1/2}=1.0\times 10^{6}(m/s)$, where $\mu_0$ is the magnetic permeability, and $\tau_{\rm A}^{*}\equiv L^{*}/V_{\rm A}^{*}=345(s)$. The coefficients $c_h^2$, $c_p^2$ in Eq.(\ref{div_eq}) are constants with values 0.04 and 0.1, respectively. $\zeta$ is a diffusion coefficient of the density which avoids strong deviations from $\rho_0$ to make the simulation more robust. In this study $\zeta$ is set to $1.0 \times 10^{-4}$. The coefficients $\nu$ and $\eta$ correspond to the viscosity and resistivity, respectively, where $\nu$ is fixed at $1.0\times 10^{-3}$ and $\eta$ depends on the NLFFF extrapolation and data-based MHD simulations. The difference between the NLFFF and the data-based MHD simulations is only in the treatment of the magnetic field at the bottom boundary. On the other boundary surfaces, the normal component is fixed in time, and the tangential components are allowed to evolve in accordance with the induction equation, in all presented simulations. The treatment of other values (velocity, density, and potential) is also the same at all the boundaries in each simulation where the velocity is fixed to zero, the density is fixed to the initial value and the normal derivative of $\phi$ is zero. This assumption drastically simplifies the numerical boundary treatment without noticeably altering the simulation results as the field strength ($|{\bf B}|$) on these five boundary surfaces is much smaller than that on the bottom boundary surface and in the bottom-center part of the simulation domain. A numerical box of 1.0 $\times$ 1.0 $\times$ 1.0 in the non-dimensional scale is divided into 320 $\times$ 320 $\times$ 320 grid points. This is proceeded by  $3 \times 3$ binning of the data points of the boundary data. 

% ---------------------------------------------------
  \subsection{Observational Data}
\begin{figure}
\plotone{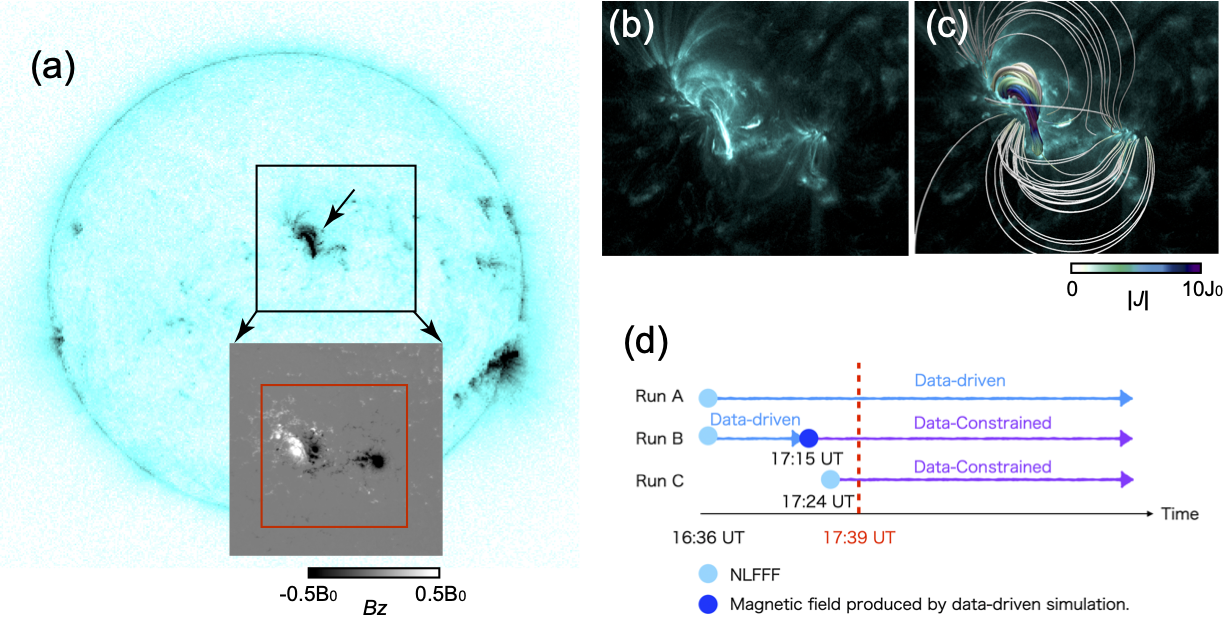}
\caption{ (a) Extreme ultra violet 131 $\AA$ image obtained from AIA  at 17:39 UT which corresponds to the onset time of an M6.5 flare is exhibited. The arrow indicates the location where the M6.5 flare occurred. The insertion shows the photospheric magnetic field at 16:36 UT which is used as the boundary condition of the simulations.The white and black colors represent the positive and negative polarities, respectively, where $B_0=0.28(T)$. (b) The AIA image of AR12371 observed at 16:36 UT. The field of view corresponds to the area enclosed by the red square in (a). (c) The 3D magnetic field lines are plotted that are extrapolated under a force-free assumption. The color of the field lines corresponds to the magnitude of the current density $|{\bf J}| = \sqrt{J_x^2 + J_y^2 + J_z^2}$. These field lines are superimposed on AIA image shown in (b). $J_0=B_0/(\mu_0 L_0)=6.5\times 10^{-3}[A]$ where $L_0=345(Mm), \mu_0(= 4\pi \times 10^{-7})$ is magnetic permeability in free space. (d) Summary of Runs A -- C. The light blue, and blue circles correspond to the NLFFF and the magnetic field produced by the data-driven simulation, respectively. The light blue arrow means that the boundary is driven by the time-varying electric field (data-driven simulation) while the purple arrow means that the electric field is turned off (data-constrained simulation).
\label{f1}}
\end{figure}

% ---------------------------------------------------
Fig.\ref{f1}(a) shows a full disk of the Sun observed by an extreme ultra violet (EUV) 131 $\AA$ image obtained from Atmospheric Imaging Assembly (\citealt{Lemen2012}) onboard Solar Dynamics Observatory (SDO: \citealt{Pesnell2012}) at 16:36. The M6.5 flare was observed at the area indicated by the black arrow.We use the photospheric magnetic field taken by Helioseismic and Magnetic Imager (HMI: \citealt{Scherrer2012}) on board {\it SDO} as the bottom boundary of the NLFFF extrapolation. The HMI photospheric magnetogram data used in this study were taken on June 22, 2015 between 16:36 UT to 20:00 UT, covering the main activity of the M6.5 flare. The field of view of such HMI data is centered on the AR 12371 and consists of 960 $\times$ 960 grid points, corresponding to $\sim$ 348 $\times$ 348 Mm$^2$. The HMI data were transformed to a local Cartesian coordinate system using the same CEA projection that is used to produce the standard space-weather HMI active region patch (SHARP) format (\citealt{Bobra2014}). The magnified insert in Fig.\ref{f1}(a) shows a snapshot of the photospheric magnetic field of the AR12371 observed at 16:36 UT that is used as the bottom boundary condition.This AR is comprised of the leading negative flux region and subsequent bipolar flux region. The sheared magnetic field lines over the magnetic polarity inversion line (PIL) of the trailing bipolar flux region is where the flare begins. Figure \ref{f1}(b) shows an AIA 131 $\AA$ image at 16:36 UT. This field of view corresponds to the area enclosed by the red square in Fig.\ref{f1}(a). We found that strong emission is observed up to one hour before the flare. 

% ------------------------------------------------------
  \subsection{Nonlinear Force-free Field Extrapolation}
% ------------------------------------------------------
We extrapolate the pre-flare NLFFF to understand the pre-eruption magnetic field and use it as the initial conditions for the data-based MHD simulations. First, we calculate the potential field as the initial condition of the NLFFF extrapolation, based on the method of \cite{Sakurai1982}. This potential field is also extrapolated from the photospheric magnetic field. Next we iterate equations(\ref{eq_motion})-(\ref{div_eq}) to obtain the NLFFF where $\eta = 5.0\times 10^{-5}+ 1.0\times 10^{-3} |{\bf J}\times{\bf B}||{\bf v}|^2/{\bf |B|^2}$. This resistivity is given for fast force-free convergence of the magnetic field above (\citealt{Inoue2011}). In this step, we assume $\rho = |{\bf B}|$ instead of solving equation (\ref{eq_mass}) to ease the relaxation of the simulation by equalizing the Alf{\'v}en speed in space. Therefore, this density has no physical significance in this stage. During the iteration, the three  components of the magnetic field are fixed at the bottom boundary. In order to avoid undesired sudden jumps near the boundary during the iterations, the velocity is adjusted as follows, ${\bf v} \Leftarrow min (1, 0.04/ M_A) {\bf v}$ where $M_A(=|{\bf v}|/|{\bf v}_A|)$ is the Alf{\' v}en Mach number. The velocity is limited to 0.04 $\times$ Alf{\' v}en Mach number or less. The details of the method are described fully in \cite{Inoue2016}. The NLFFF extrapolations are applied to the photospheric magnetic field observed at 16:36 UT (Fig.\ref{f1}(a)) and 17:24 UT, about 1 hour, and 15 minutes before the M6.5 flare (17:39 UT), respectively. The NLFFF at 16:36 UT is shown in Fig.\ref{f1}(c), which captures the field lines well, as inferred from EUV image. The strong current density region corresponds to the strong EUV emission region.

% --------------------------------------------
  \subsection{Data-based MHD Simulations}
% --------------------------------------------
The data-driven MHD simulation employs equations (\ref{eq_mass}) - (\ref{div_eq}) where the bottom magnetic fields are driven by the electric fields {(${\bf E}$)}. The resistivity $\eta$ has value $1.0\times 10^{-5}$. The electric fields are derived from the method proposed by \cite{Hayashi2018} and \cite{Hayashi2019}, and the numerical method developed in \cite{Inoue2022} is applied. The electric fields are obtained through the three Poisson equations, which are derived from an induction equation that includes the time derivative of the photospheric magnetic field as the source term. Since the cadence of our boundary data is 12 minutes ($\approx 2.1\tau_A$ in this simulation), we calculate the electric fields in advance from the photospheric magnetic field using observed times that frame the current simulation time step and the coronal magnetic field is driven as $\partial_t{\bf B}=-\nabla\times{\bf E}$ with the same electric field in 12 minutes and then switch to the next electric field. Note that the magnetic field is driven at the location where the strength of the magnetic field ($B_t=\sqrt{bx^2+by^2+bz^2})$ at the bottom satisfies more than $0.05$ ($\approx 140(G)$). That is, the extremely weak magnetic field (outside the active region) is fixed in time with initial value (NLFFF). We repeat this process. The detailed processes summarised here are described further in Section 2.3 in \cite{Inoue2022}. The initial density $\rho$ is given as $\rho ({\bf r}) = \rho_B \exp(-z/H_s)$, where $\rho_B$ is density at the bottom boundary given as 1 which corresponds to $\rho^{*}$. $H_s$ is a scale height given as 0.1 which corresponds to $3.456\times 10^{7}(m)$. 

We ran three different data-based MHD simulations, Run A to Run C. Run A is a full data-driven simulation where the initial condition is the NLFFF extrapolated at 16:36 UT, and a time varying electric field is imposed on the bottom.  Run B was also data-driven simulation, carried out using data by 17:15 UT, corresponding to 24 minutes before the flare. The driving electric field is set to zero, so that the bottom boundary $B_x$ and $B_y$ are allowed to evolve in accordance with the induction equation (Eq. \ref{induc_eq}), while $B_z$ is fixed with time. Note that the velocity and the density profile are reset to $|{\bf v}|=0$ and $\rho ({\bf r}) = \rho_B \exp(-z/H_s)$ at 17:15 UT in Run B to avoid a tight time step in the simulation. This corresponds to a data-constrained simulation done by \cite{Inoue2018}. Run C was a data-constrained simulation where the NLFFF at 17:24 UT is used as the initial condition. The initial conditions of the other physical values and the boundary conditions are the same as Run B after $t=$17:15 UT. These simulation steps are summarized in Fig.\ref{f1}(d).

% =======================================================================================================================================================
    \section{Results} \label{sec:floats}
% =======================================================================================================================================================
  \subsection{Pre-eruption Magnetic Field Produced by the NLFFF Extrapolation and Data-driven Simulation}
Figure \ref{f2} shows the magnetic field before the flare obtained from the NLFFF extrapolation and data-driven MHD simulation. The vertical cross section plots $|{\bf J}|/|{\bf B}|$ distribution. Panel (a) shows the magnetic field extrapolated from the photospheric magnetic field observed at 16:36 UT, while panels (b) and (c) show show the magnetic fields extrapolated at 17:24 UT. The magnetic field structures are seen to be almost identical. The lower panels (d-f) show the result of the data-driven MHD simulation at 16:36 UT and 17:15 UT. Panels (c) and (f) show the top-down views of the NLFFF at 17:24 UT, and the data-driven magnetic field at 17:15 UT respectively.

The data-driven simulation starts with the NLFFF at 16:36 UT shown in Fig.\ref{f2} (d). The electric field given at the bottom boundary then drives it for 40 minutes, eventually producing the magnetic field as shown in Fig. \ref{f2} (e) and (f). This magnetic field structure shown in panels (e) and (f) is clearly different from the NLFFF shown in panels (b) and (c). In particular, the simulation has highly twisted field lines connecting to distant sunspots, which is consistent with the analysis by \cite{Kliem2021} and a data-driven simulation done by \cite{He2020}. A current sheet structure is formed below the long twisted field lines in Fig. \ref{f2} (e) while the NLFFF shows that the elbow-shaped sheared field lines  only lie close to the surface at 17:24 UT (Fig. \ref{f2} (b)). From the supplemental movie of the data-driven simulation, the highly twisted field lines are created through magnetic reconnection between the sheared field lines formed in the NLFFF at 16:36 UT. These are then lifted up, forming the current sheet structure. This is a typical formation process of a pre-eruption magnetic flux rope ({\it e.g.}, \citealt{Amari2000}, \citealt{Jiang2014}). Our results show that the magnetic structure generated by the NLFFF and  the data-driven simulation are completely different for this AR. 

% --------------------------------------------------------------------------------------
  \subsection{Eruption Dynamics in the Data-based Simulations}
\begin{figure}
\plotone{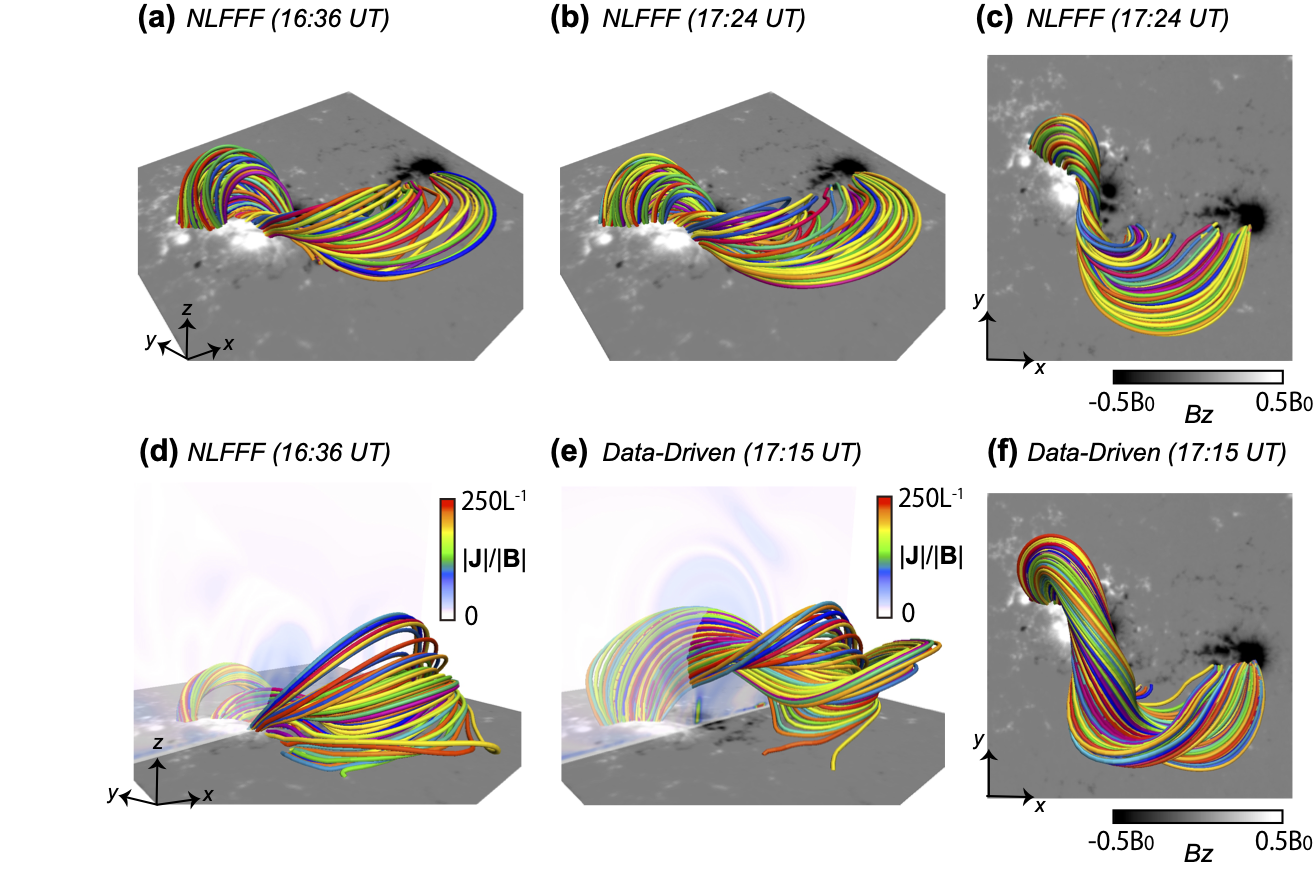}
\caption{ (a)-(b) The magnetic fields (NLFFF) that are extrapolated from the photospheric magnetic fields obtained at 16:36 UT and 17:24 UT. (c) The top view of the NLFFF at 17:24 UT. (d)-(e) The temporal evolution of the magnetic field lines during the data-driven MHD simulation. The vertical cross section plots ${|\bf J}|/|{\bf B}|$. (f) shows the top view of (e).$B_0=0.28(T)$ and $L^{-1}=1/(\mu_0 L_0)=2.3\times 10^{-3}[1/m]$ where $L_0=345(Mm), \mu_0(= 4\pi \times 10^{-7})$ is magnetic permeability in free space.
\label{f2}}
\end{figure}

\begin{figure}
\plotone{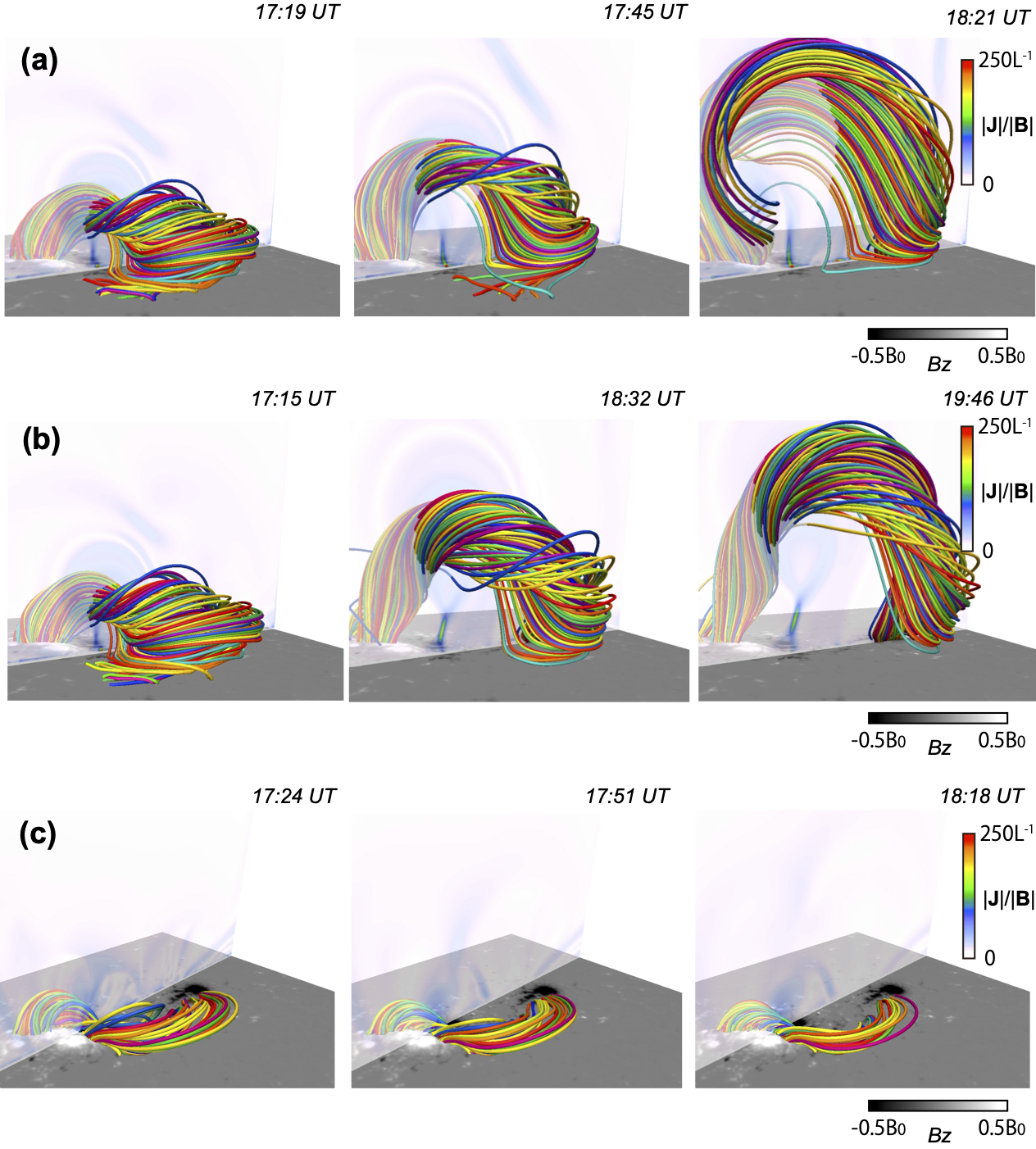}
\caption{The temporal evolution of the magnetic field lines obtained from the data-based MHD simulations from Run A to Run C. The coloured lines correspond to the field lines and $|{\bf J}|/|{\bf B}|$ is plotted on the vertical cross section. $B_0=0.28(T)$ and $L^{-1}=1/(\mu_0 L_0)=2.3\times 10^{-3}[1/m]$ where $L_0=345(Mm), \mu_0(= 4\pi \times 10^{-7})$ is magnetic permeability in free space. (a) The result of Run A: the magnetic field obtained from the full data-driven simulation. (b) The result of Run B: the data-driven MHD simulation is conducted by 17:15 UT as in (a) and then the data-constrained simulation is carried out. (c) The result of Run C: the NLFFF, which is extrapolated at 17:24 UT, is used as the initial condition for the data-constrained simulation. This calculation employs the same bottom boundary condition as Run B after t = 17:15 UT. An animation of the temporal evolution is available in the online Journal. The animation proceeds from $t=16:36$ UT to $t=18:32$ UT for Run A,  $t=17:15$ UT to $t=20:50$ UT for Run B, and $t=17:24$ UT to $t=18:18$ UT for Run C.
\label{f3}}
\end{figure}

% --------------------------------------------------------------------------------------
We show the simulation results for Runs A-C. Figure \ref{f3}(a) shows the temporal evolution of the magnetic field lines and $|{\bf J}|/|{\bf B}|$ distribution for a vertical cross section in the full data-driven simulation (Run A). In this case, the magnetic field lines erupt and a current sheet is formed under the erupting magnetic field lines. These results are consistent with \cite{Liu2019}. Figure \ref{f3}(b) shows the results of Run B, where the electric field given at the bottom surface is turned off at 17:15 UT. This simulation also shows the eruption even though the photospheric driving is stopped 30 minutes before the flare. The magnetic field is similar to that in Run A until the middle of the eruption when their structures differ. Furthermore, the velocity of the ascending eruption in Run B is slower than in the case of Run A. Figure \ref{f3}(c) shows the results of Run C where the NLFFF at 17:24 UT is employed as the initial condition and the bottom electric field is switched off, which is same condition as Run B, after t = 17:15 UT. Although the numerical residual force works on the NLFFF, the magnetic field changes little throughout the simulation. We therefore found differing stability between the magnetic fields produced by the NLFFF extrapolation and the data-driven simulation. From these results, we found that the magnetic field shown in Fig.\ref{f2}(e)  produced by the data-driven simulation can lead to the eruption even in data-constrained simulation, while the NLFFF (Fig.\ref{f2}(b)) cannot produce an eruption. It therefore seems that the data-driven simulation is preferable for reproducing the pre-eruption magnetic field compared to the NLFFF extrapolation. It is important to note however,that such results depend on the active region in question. For instance, \cite{Guo2019} reported that the pre-eruption magnetic fields for another active region are almost same in both the NLFFF extrapolation and in the data-driven simulation.

\begin{figure}
\plotone{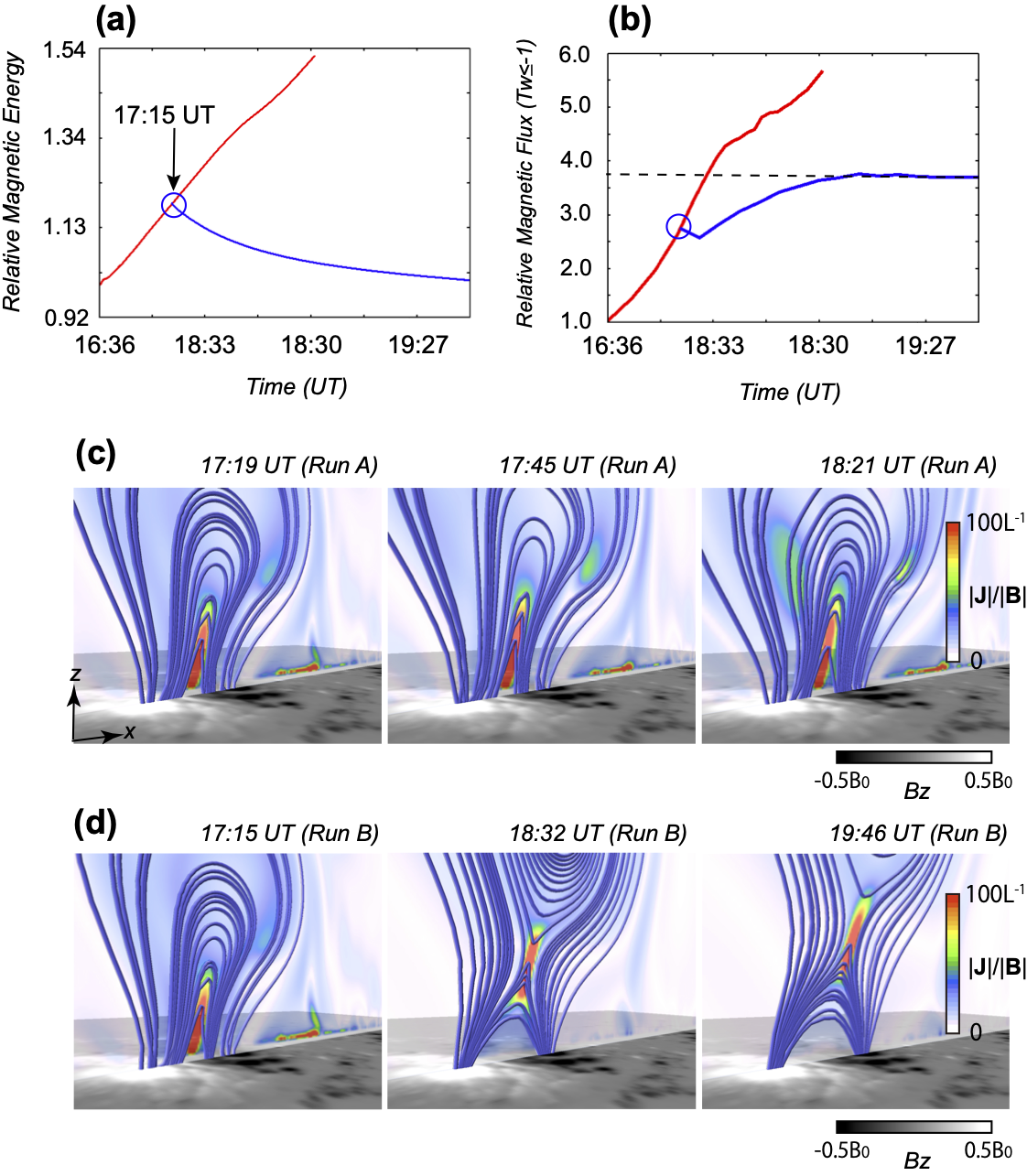}
\caption{(a)The temporal evolution of the magnetic energy obtained from Run A (red) and Run B (blue). The vertical value is normalized by the initial value at $t=0$ in the horizontal axis that corresponds to the NLFFF extrapolated at 16:36 UT. (b) The temporal evolution of the magnetic flux that satisfies $T_w \le -1.0$. The format is the same as in (a). The vertical axis is normalized by the initial value that corresponds to $1.4 \times 10^{21}[{\rm Mx}]$. The dashed horizontal line sets the value at which the magnetic flux in Run B saturates. (c-d) The magnetic field lines projected in the $x-z$ plane are plotted for Runs A and Run B where $B_0=0.28(T)$ and $L^{-1}=1/(\mu_0 L_0)=2.3\times 10^{-3}[1/m]$ where $L_0=345(Mm), \mu_0(= 4\pi \times 10^{-7})$ is magnetic permeability in free space.
\label{f4}}
\end{figure}

 To see the differences between Run A and Run B, especially after the eruption, we plotted the temporal evolution of the magnetic energy of Run A (red) and Run B (blue) in Fig.\ref{f4}(a). Note that the electric field in Run B was turned off at the time indicated by blue circle. The subsequent results were quite different. The magnetic energy in Run A increased with time, even after the flare, while the energy in Run B decreased monotonically. The Fig.\ref{f4}(b) shows the temporal evolution of the magnetic flux where the twist (\citealt{Berger2006}) is defined as 
\begin{equation}
    T_w = \frac{1}{4\pi}\int\frac{{\bf B}\cdot{\bf \nabla}\times {\bf B}}{|{\bf B}|^2}dl,
\end{equation}
and satisfies the condition of $T_w \le -1.0$. We calculate $T_w$ for each field line and determine the footpoints of the field lines with $T_w \le -1.0$ and then we calculate the magnetic flux. The magnetic flux in Run B (blue) saturates during the evolution because the sheared magnetic field lines present in the active region (supplied as the twisted magnetic field lines to the erupting magnetic flux rope through a reconnection) are depleted and the magnetic flux accumulated in the magnetic flux rope is limited (\citealt{Inoue2018}). On the other hand, the magnetic flux in Run A (red) continuously increases with time. Therefore, the erupting field lines are continuously driven unlike in Run B. Figures \ref{f4}(c) and (d) exhibit a temporal evolution of field lines after the eruption, which are mapped on the x-z plane with the $|{\bf J}|/|{\bf B}|$ distribution, for Run A and Run B, respectively. The location of the strong current region in Run A is almost fixed with time and the field lines are stretched continuously ({\it e.g.},\citealt{Mikic1988}) even after the eruption, {\i.e.}, more sheared field lines are created in the current sheet, which would encourage acceleration of the eruption. In contrast to Run A, Run B shows that the strong current region gets higher with time and the post-flare loops are formed associated with reconnection (\citealt{Shibata1996}). \cite{Jing2016} and \cite{Wang2017} reported flare-ribbons in this event that clearly separate from each other. Therefore, this result suggests that Run B is more consistent with the observation.

% --------------------------------------------------------------------------------------------
% Figure 1
% --------------------------------------------------------------------------------------------

% --------------------------------------------------------------------------------------------
% Figure 2
% --------------------------------------------------------------------------------------------

% --------------------------------------------------------------------------------------------
% Figure 3
% --------------------------------------------------------------------------------------------

% --------------------------------------------------------------------------------------------
% Figure 4
% --------------------------------------------------------------------------------------------

% ================================================================================================================================================
    \section{Discussion}
\begin{figure}
\plotone{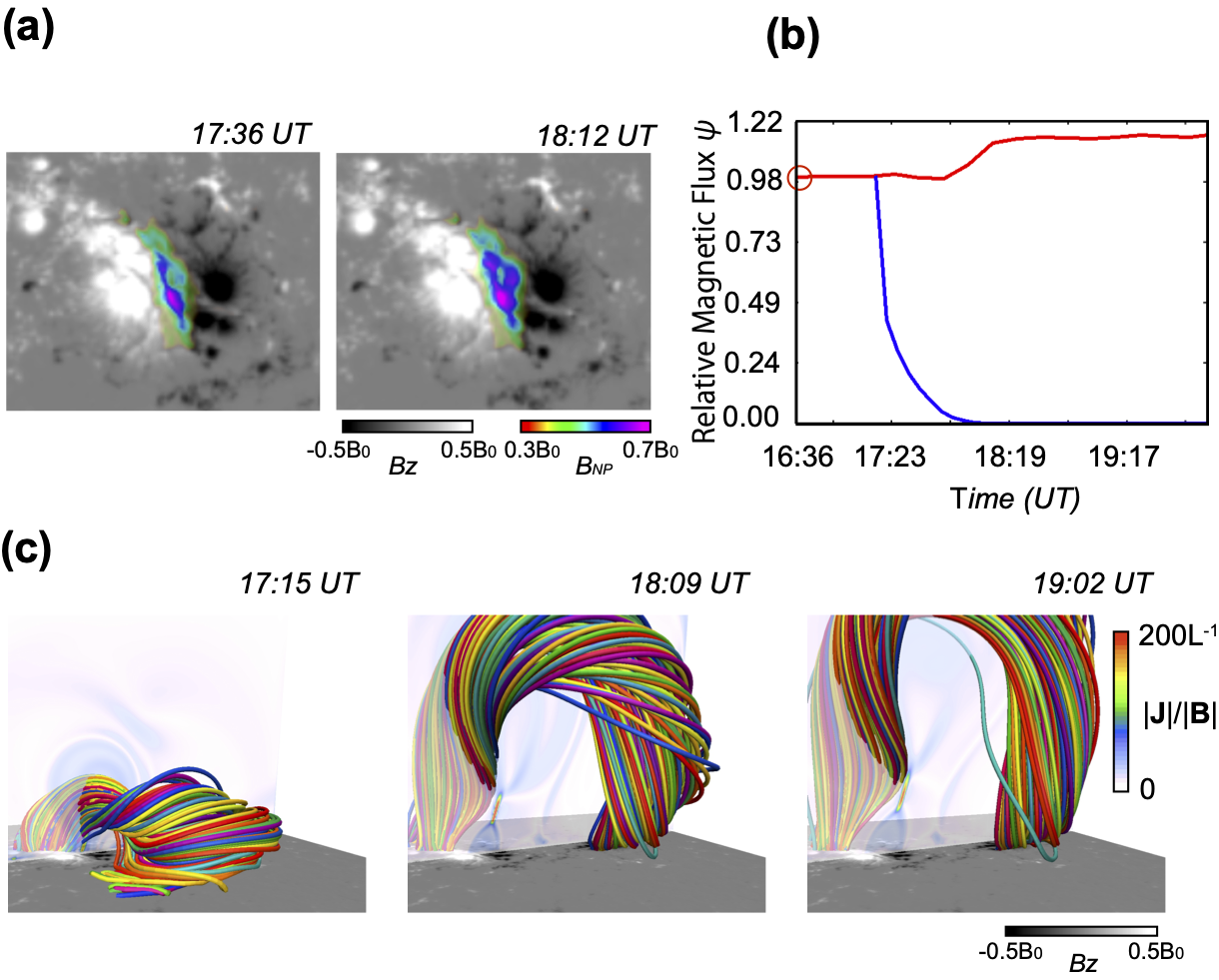}
\caption{(a) The distribution $B_{NP}$ that is defined as $B_{NP} = {|\bf B}-{\bf B_p}|$ calculated on the photosphere is plotted in a value range from 0.3$B_{0}$  to 0.7$B_{0}$ at 17:36 UT and 18:12 UT, respectively, where ${\bf B_p}$ is the potential field and $B_{0}=0.28(T)$. (b) Temporal evolution of $\psi$, which is defined as $\psi=\int B_{NP} dS $ on the photosphere, is plotted in the specific range from $B_{NP} \ge 0.3$ in red for the observed magnetic field and in blue for Run B, respectively. The value on the vertical axis is normalized by the value ($6.7 \times 10^{21}[{\rm Mx}])$ at 16:36 UT (NLFFF) that is indicated by the red circle. (c) The 3D magnetic field lines obtained from Run B+. The format is the same as Fig.\ref{f3}. An animation of the temporal evolution for (c) is available in the online Journal. The animation proceeds from $t=17:15$ UT to $t=19:02$ UT.
\label{f5}}
\end{figure}

% ================================================================================================================================================
   We discuss the differences in Runs A and B as seen in Figs. \ref{f3} and \ref{f4}. The magnetic structures are obviously different after the eruption in both. These would be caused by the different boundary conditions, {\it i.e.}, whether the electric field is given or not at the bottom surface after the eruption. We revisit the photospheric magnetic field that is the origin of the electric field. The two sub-panels of Figure \ref{f5}(a) show the spatial distribution of the absolute non-potential component accumulated in the observed photospheric magnetic field at two instants. The non-potential component is defined as ${B_{NP}} = |{\bf B} - {\bf B_P}|$ and measured at the photosphere, where ${\bf B_p}$ is the potential field. Note that values displayed are in the range 0.3 to 0.7, which are concentrated on the PIL.  The M6.5 flare started at 17:39 UT, following which $B_{NP}$ is enhanced during the flare. Figure {\ref{f5}}(b) plots the temporal evolution of $\psi$, where $\psi$ is defined as $\psi=\int B_{NP} dS $, which focuses on the specific range $B_{NP} \ge 0.3$. The results from the observed magnetic field and Run B are plotted in red and blue, respectively. We found that $\psi$ in the observed magnetic field suddenly increases during the flare and keeps the high non-potential component. Conversely the evolution in Run B steeply decreases toward the potential field. The enhancement shown in the observation has been suggested as the back reaction of the flaring process (\citealt{Wang1994}, \citealt{Liu2012}). The data-driven simulation is simply giving a time-dependent boundary value and therefore cannot distinguish between whether the information is derived from photospheric motion or flare cause like the back reaction, as discussed in \cite{Hayashi2018}.  Therefore, the data-driven simulation requires appropriate handling of information that comes from above, after the flare. 

We found that Run B reproduces the phenomena associated with flares, such as post-flare loops, well.  Run A on the other hand does not. These results suggest that a relaxation process is required at the bottom boundary to produce those phenomena (\citealt{Inoue2015}). However, the relaxation process in Run B is inconsistent with the evolution of the photospheric magnetic field. This inconsistency arises from the difference between the timescale of the observed photosphere, $\tau^{p}$ and that of the simulated photosphere, $\tau^{sp}$. Since the photosphere is composed of denser plasma than the corona, $\tau^{p}$ is much longer than the typical coronal Alf{\' v}en timescale, $\tau_A^{c}$. In contrast, $\tau^{sp}$ is mostly determined by $\tau_A^{c}$ in Run B because the dense plasma is not taken into account in the photosphere. Thus, $\tau^{sp}$ is much shorter than $\tau^{p}$. Therefore, a contradiction arises in Run B such as the photospheric horizontal magnetic fields immediately get close to the potential field on $\tau_A^C$. 

Finally we note the velocity of the rising flux rope in Run B because it is very slow. One reason for this is that the viscosity in the simulation is set to a large value to make the calculation more robust. This is also the case in Run A. The other reason is because of the density distribution. Additionally, we calculated Run B+ in which the density is given $\rho({\bf r}, t)=|{\bf B}({\bf r}, t)|$ in Run B. The 3D field lines are shown in Fig.\ref{f5}(c) and are seen to be almost the same as Run B. However, the eruption progressed much faster than in the case of Run B. Therefore, the modeling of the rising velocity is much better reproduced by the density model.

From these results, more realistic modeling, in particular, the proper treatment of stratification to account for the much larger mass density, is important to improve the data-driven simulation. In addition, it would be helpful if we could use the time varying magnetic fields, which include the magnetic field variation on the coronal time scale $\tau_A^C$, as the boundary condition as zero-beta assumption is applicable. In other words, the magnetic field used should be from higher in the atmosphere than the photospheric magnetic field, for example, the chromospheric magnetic field (\citealt{Kawabata2020}). The use of the chromospheric magnetic field would be a key issue not only for NLFFF extrapolation as shown in \cite{Fleishman2019}, but also a data-driven simulation as discussed in \cite{Jiang2020}. Note that chromospheric magnetic field is also contaminated after the flare and it may be usable only before the flare. However, the ${\bf \nabla} \cdot {\bf B}=0$ issue is not resolved through the use of chromospheric observations as it is not satisfied within the 2 dimensional plane. As \cite{Hayashi2018} suggested, the vertical gradients of ${\bf B}$ might solve this problem, therefore, simultaneous observation of 2 layers of the solar atmosphere in the vertical direction would be required. Our results suggest that the improvement of both methods and techniques of the magnetic field observations are important to further improve the reliability of data-based MHD simulations.

% --------------------------------------------------------------------------------------------
% Figure 5
% --------------------------------------------------------------------------------------------
    
% ================================================================================================================================================
    \acknowledgments
% ================================================================================================================================================
\begin{acknowledgments}
We thank to referee for useful comments. We are grateful to Dr. Magnus Woods for reading this manuscript. This work was supported by the National Science Foundation  under grant AGS-1954737,  AGS-2145253, AGS-2149748 and AST-2204384, and National Aeronautics and Space Administration under grants 80NSSC21K1671 80NSSC21K0003 and 80HQTR20T0067. TM is supported by JSPS KAKENHI Grant Numbers JP20K11851, JP20H00156. The visualization was done by VAPOR (\citealt{Clyne2005}, \citealt{Clyne2007}).   
\end{acknowledgments}




\begin{thebibliography}{}
\bibitem[Amari et al.(2000)]{Amari2000} Amari, T., Luciani, J.~F., Mikic, Z., et al.\ 2000, \apjl, 529, L49. doi:10.1086/312444
\bibitem[Berger \& Prior(2006)]{Berger2006} Berger, M.~A. \& Prior, C.\ 2006, Journal of Physics A Mathematical General, 39, 8321. doi:10.1088/0305-4470/39/26/005
\bibitem[Bobra et al.(2014)]{Bobra2014} Bobra, M.~G., Sun, X., Hoeksema, J.~T., et al.\ 2014, \solphys, 289, 3549. doi:10.1007/s11207-014-0529-3
\bibitem[Cheung \& DeRosa(2012)]{Cheung2012} Cheung, M.~C.~M. \& DeRosa, M.~L.\ 2012, \apj, 757, 147. doi:10.1088/0004-637X/757/2/147
\bibitem[Clyne \& Rast(2005)]{Clyne2005} Clyne, J., \& Rast, M.\ 2005, \procspie, 284
\bibitem[Clyne et al.(2007)]{Clyne2007} Clyne, J., Mininni, P., Norton, A., et al.\ 2007, New Journal of Physics, 9, 301
\bibitem[Dacie et al.(2018)]{2018ApJ...862..117D} Dacie, S., T{\"o}r{\"o}k, T., D{\'e}moulin, P., et al.\ 2018, \apj, 862, 117. doi:10.3847/1538-4357/aacce3
\bibitem[Dedner et al.(2002)]{Dedner2002} Dedner, A., Kemm, F., Kr{\"o}ner, D., et al.\ 2002, Journal of Computational Physics, 175, 645. doi:10.1006/jcph.2001.6961
\bibitem[Fleishman et al.(2019)]{Fleishman2019} Fleishman, G., Mysh'yakov, I., Stupishin, A., et al.\ 2019, \apj, 870, 101. doi:10.3847/1538-4357/aaf384
\bibitem[Guo et al.(2019)]{Guo2019} Guo, Y., Xia, C., Keppens, R., et al.\ 2019, \apjl, 870, L21. doi:10.3847/2041-8213/aafabf
\bibitem[Hayashi et al.(2018)]{Hayashi2018} Hayashi, K., Feng, X., Xiong, M., et al.\ 2018, \apj, 855, 11. doi:10.3847/1538-4357/aaacd8
\bibitem[Hayashi et al.(2019)]{Hayashi2019} Hayashi, K., Feng, X., Xiong, M., et al.\ 2019, \apjl, 871, L28. doi:10.3847/2041-8213/aaffcf
\bibitem[He et al.(2020)]{He2020} He, W., Jiang, C., Zou, P., et al.\ 2020, \apj, 892, 9. doi:10.3847/1538-4357/ab75ab
\bibitem[Inoue et al.(2014)]{Inoue2014} Inoue, S., Hayashi, K., Magara, T., et al.\ 2014, \apj, 788, 182
\bibitem[Inoue et al.(2015)]{Inoue2015} Inoue, S., Hayashi, K., Magara, T., et al.\ 2015, \apj, 803, 73. doi:10.1088/0004-637X/803/2/73
\bibitem[Inoue et al.(2022)]{Inoue2022} Inoue, S., Hayashi, K., \& Miyoshi, T.\ 2022, arXiv:2210.07492. doi:10.48550/arXiv.2210.07492
\bibitem[Inoue(2016)]{Inoue2016} Inoue, S.\ 2016, Progress in Earth and Planetary Science, 3, 19
\bibitem[Inoue et al.(2018)]{Inoue2018} Inoue, S., Kusano, K., B{\"u}chner, J., et al.\ 2018, Nature Communications, 9, 174
\bibitem[Inoue et al.(2011)]{Inoue2011} Inoue, S., Kusano, K., Magara, T., et al.\ 2011, \apj, 738, 161. doi:10.1088/0004-637X/738/2/161
\bibitem[Jiang et al.(2022)]{Jiang2022} Jiang, C., Feng, X., Guo, Y., et al.\ 2022, The Innovation, 3, 100236. doi:10.1016/j.xinn.2022.100236
\bibitem[Jiang et al.(2013)]{Jiang2013} Jiang, C., Feng, X., Wu, S.~T., et al.\ 2013, \apjl, 771, L30. doi:10.1088/2041-8205/771/2/L30
\bibitem[Jiang \& Toriumi(2020)]{Jiang2020} Jiang, C. \& Toriumi, S.\ 2020, \apj, 903, 11. doi:10.3847/1538-4357/abb5ac
\bibitem[Jiang et al.(2014)]{Jiang2014} Jiang, C., Wu, S.~T., Feng, X., et al.\ 2014, \apj, 780, 55. doi:10.1088/0004-637X/780/1/55
\bibitem[Jiang et al.(2016)]{Jiang2016} Jiang, C., Wu, S.~T., Feng, X., et al.\ 2016, Nature Communications, 7, 11522. doi:10.1038/ncomms11522
\bibitem[Jing et al.(2016)]{Jing2016} Jing, J., Xu, Y., Cao, W., et al.\ 2016, Scientific Reports, 6, 24319. doi:10.1038/srep24319
\bibitem[Kaneko et al.(2021)]{Kaneko2021} Kaneko, T., Park, S.-H., \& Kusano, K.\ 2021, \apj, 909, 155. doi:10.3847/1538-4357/abe414
\bibitem[Kang et al.(2019)]{Kang2019} Kang, J., Inoue, S., Kusano, K., et al.\ 2019, \apj, 887, 263
\bibitem[Kawabata et al.(2020)]{Kawabata2020} Kawabata, Y., Asensio Ramos, A., Inoue, S., et al.\ 2020, \apj, 898, 32. doi:10.3847/1538-4357/ab9816
\bibitem[Kilpua et al.(2021)]{Kilpua2021} Kilpua, E.~K.~J., Pomoell, J., Price, D., et al.\ 2021, Frontiers in Astronomy and Space Sciences, 8, 35. doi:10.3389/fspas.2021.631582
\bibitem[Kliem et al.(2021)]{Kliem2021} Kliem, B., Lee, J., Liu, R., et al.\ 2021, \apj, 909, 91. doi:10.3847/1538-4357/abda37
\bibitem[Leake et al.(2017)]{Leake2017} Leake, J.~E., Linton, M.~G., \& Schuck, P.~W.\ 2017, \apj, 838, 113. doi:10.3847/1538-4357/aa6578
\bibitem[Lemen et al.(2012)]{Lemen2012} Lemen, J.~R., Title, A.~M., Akin, D.~J., et al.\ 2012, \solphys, 275, 17. doi:10.1007/s11207-011-9776-8
\bibitem[Liu et al.(2019)]{Liu2019} Liu, C., Chen, T., \& Zhao, X.\ 2019, \aap, 626, A91. doi:10.1051/0004-6361/201935225
\bibitem[Liu et al.(2012)]{Liu2012} Liu, C., Deng, N., Liu, R., et al.\ 2012, \aas
\bibitem[Mikic et al.(1988)]{Mikic1988} Mikic, Z., Barnes, D.~C., \& Schnack, D.~D.\ 1988, \apj, 328, 830. doi:10.1086/166341
\bibitem[Muhamad et al.(2017)]{Muhamad2017} Muhamad, J., Kusano, K., Inoue, S., et al.\ 2017, \apj, 842, 86. doi:10.3847/1538-4357/aa750e
\bibitem[Pesnell et al.(2012)]{Pesnell2012} Pesnell, W.~D., Thompson, B.~J., \& Chamberlin, P.~C.\ 2012, \solphys, 275, 3. doi:10.1007/s11207-011-9841-3
\bibitem[Sakurai(1982)]{Sakurai1982} Sakurai, T.\ 1982, \solphys, 76, 301. doi:10.1007/BF00170988
\bibitem[Scherrer et al.(2012)]{Scherrer2012} Scherrer, P.~H., Schou, J., Bush, R.~I., et al.\ 2012, \solphys, 275, 207. doi:10.1007/s11207-011-9834-2
\bibitem[Schrijver et al.(2008)]{Schrijver2008} Schrijver, C.~J., DeRosa, M.~L., Metcalf, T., et al.\ 2008, \apj, 675, 1637. doi:10.1086/527413
\bibitem[Shibata(1996)]{Shibata1996} Shibata, K.\ 1996, Advances in Space Research, 17, 9. doi:10.1016/0273-1177(95)00534-L
\bibitem[Shibata \& Magara(2011)]{Shibata2011} Shibata, K. \& Magara, T.\ 2011, Living Reviews in Solar Physics, 8, 6. doi:10.12942/lrsp-2011-6
\bibitem[Sun et al.(2012)]{Sun2012} Sun, X., Hoeksema, J.~T., Liu, Y., et al.\ 2012, \apj, 748, 77. doi:10.1088/0004-637X/748/2/77
\bibitem[Toriumi et al.(2020)]{Toriumi2020} Toriumi, S., Takasao, S., Cheung, M.~C.~M., et al.\ 2020, \apj, 890, 103. doi:10.3847/1538-4357/ab6b1f
\bibitem[Toriumi, \& Wang(2019)]{Toriumi2019} Toriumi, S., \& Wang, H.\ 2019, Living Reviews in Solar Physics, 16, 3
\bibitem[Wang et al.(1994)]{Wang1994} Wang, H., Ewell, M.~W., Zirin, H., et al.\ 1994, \apj, 424, 436. doi:10.1086/173901
\bibitem[Wang et al.(2017)]{Wang2017} Wang, H., Liu, C., Ahn, K., et al.\ 2017, Nature Astronomy, 1, 0085. doi:10.1038/s41550-017-0085
\end{thebibliography}
\end{document}